\pdfoutput=1

\documentclass[%
 reprint,
 amsmath, amssymb,
 aps, prl
]{revtex4-1}

\usepackage{graphicx}
\usepackage{dcolumn}
\usepackage{bm}
\usepackage{bbm}
\usepackage{physics}
\usepackage{siunitx}
\usepackage{floatrow}
\usepackage{hyperref}

\begin{document}

\title{Long-distance entangling gates between quantum dot spins mediated by a superconducting resonator}

\author{Ada Warren}
\author{Edwin Barnes}
\author{Sophia E. Economou}
\affiliation{Department of Physics, Virginia Tech, Blacksburg VA 24061, USA}

\begin{abstract}
Recent experiments with silicon qubits demonstrated strong coupling of a microwave resonator to the spin of a single electron in a double quantum dot, opening up the possibility of long-range spin-spin interactions. We present our theoretical calculation of effective interactions between distant quantum dot spins coupled by a resonator, and propose a protocol for fast, high-fidelity two-qubit gates consistent with experimentally demonstrated capabilities. Our simulations show that, in the presence of noise, spin-spin entangling gates significantly outperform cavity-mediated gates on charge qubits.
\end{abstract}
\maketitle

Solid-state electronic spins are promising candidates for quantum information processing \cite{Kawakami16}. Electronic spins in isotopically-purified silicon have been shown to exhibit long coherence times \cite{Veldhorst14}, and mature silicon fabrication technologies improve prospects of scalable, low-cost silicon-based quantum computing technologies \cite{Zwanenburg13}. Much research on quantum computing with spins has focused on achieving entanglement with fermionic exchange or dipole-dipole interactions \cite{Loss98, Dehollain16}. These interactions are short-range, creating significant challenges toward achieving long-range entanglement necessary for scalable quantum processors \cite{Ladd10}.

One proposed solution is the introduction of a superconducting microwave resonator, as in circuit QED. Coupling between resonator modes and the electronic spins gives rise to a long-distance, effective spin-spin coupling mediated by cavity photons \cite{Petersson12, Blais07}. Unfortunately, such approaches suffer from the very weak ($<\SI{1}{\kilo\hertz}$) magnetic dipole coupling between electronic spins and radiation modes, making realization of strong spin-photon coupling challenging \cite{Schoelkopf08, Amsuss11, Schuster10}.

\begin{figure}
\includegraphics[width=\textwidth]{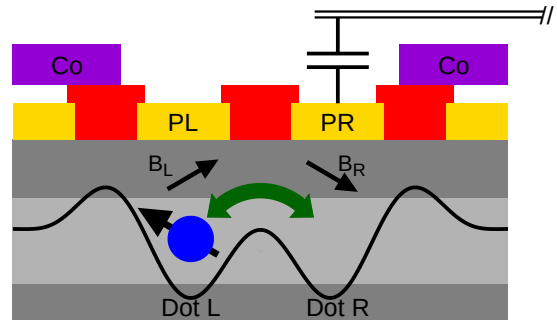}
\caption{Diagram of a DQD system. An electron in a Si/SiGe heterostructure is trapped in the double-well potential and allowed to tunnel between the two dots. The double-well potential is defined by aluminum gates on top of the heterostructure, shown in red and gold. The plunger gate PR above the right dot is capcitatively coupled to a probe in the microwave resonator. Nearby Co micromagnets create different magnetic fields $\vec{B_L}$ and $\vec{B_R}$ at the left and right dots, respectively.}
\label{fig:DQDsys}
\end{figure}

Recent breakthrough experimental work has demonstrated a coherent interface between electronic spins in semiconductors and microwave-frequency photons in a superconducting resonator \cite{Viennot15, Mi18, Samkharadze18, Landig18}. In the scheme of Refs. \cite{Mi18, Samkharadze18}, a single excess electron is trapped in a gate-defined silicon double quantum dot (DQD) near a cobalt micromagnet (see Fig. \ref{fig:DQDsys}). The large, inhomogeneous magnetic field produced by the micromagnet creates a large coupling between the electron's spin and orbital degrees of freedom \cite{Kawakami14}. A plunger gate above one dot is connected to a probe in a high-Q superconducting microwave resonator. The large electric dipole moment of the electron in the DQD system leads to strong coupling between resonator photons and the electron's orbital state \cite{Mi17, Stockklauser17, Woerkom18}. The combination of spin-orbital and orbital-photon couplings can result in a large ($\approx\SI{10}{\mega\hertz}$) effective spin-photon coupling \cite{Cottet10, Hu12}, nearly five orders of magnitude larger than typical magnetic dipole couplings \cite{Mi18}. This coupling strength improvement, along with a significant reduction in charge noise from careful gate design \cite{Mi17}, places this spin-photon system firmly into the strong coupling regime, with the effective coupling greater than both the spin decoherence rate and photon loss rate.

In addition to a greatly enhanced spin-photon coupling and reduced charge noise susceptibility, this architecture allows for dispersive spin readout and coherent spin control via driven electric dipole spin resonance (EDSR) techniques \cite{Mi18, Petersson12, Wu18}. These single-spin EDSR rotations along with phase gates and an entangling two-qubit operation compose a universal set of quantum gates \cite{DiVincenzo00}. It is natural, therefore, to ask what multi-spin operations can be achieved when these DQD systems are connected to a common resonator in a quantum bus topology \cite{Woerkom18}.

We address this timely topic by designing a fast, electrically controllable, high-fidelity entangling gate acting on silicon spin qubits. We start by presenting the effective cavity-mediated spin-spin interactions which occur in the low-energy limit of such a system. We then present numerical results supporting these theoretical findings in the case of two electronic spins and demonstrate robust electrical generation of entangling gates. For comparison, we also simulate entangling gates between qubits based on the electrons' orbital degrees of freedom. We find that spin-spin gates perform substantially better in the presence of charge noise.

To investigate multi-spin interactions, we begin with a single electron in a single DQD system coupled to the microwave resonator, as shown in Fig. \ref{fig:DQDsys}. We assume excited single dot states are sufficiently high in energy that they can be ignored. The orbital Hilbert space is then spanned by the $\ket{L}$ and $\ket{R}$ orbital states, localized to the left and right quantum dots, respectively. Following Ref. \cite{Mi17}, we further assume that the quantized resonator field affects only the average energy of the $\ket{R}$ orbital state, and that only one resonator mode, with frequency $\omega_r$, appreciably affects DQD dynamics.

Because the DQD sits in the micromagnet's large, inhomogeneous magnetic field, the spin feels different fields in the left and right dots. We take the average magnetic field $\vec{B} = (\vec{B_L} + \vec{B_R})/2$ to be along the $z$-axis and the difference in magnetic fields $\Delta\vec{B} = \vec{B_L} - \vec{B_R}$ to be in the $xz$-plane. $\vec{B_L}$ and $\vec{B_R}$ include both magnetic fields generated by the micromagnet and any externally applied field. The Hamiltonian for $N$ DQDs coupled to a resonator is \cite{Beaudoin16}
\begin{align}
\label{eq:baremultispin}
&H_N = H_r + \sum_{i = 1}^N \qty(H_{DQD i} + H_{z i} + H_{SO i} + H_{RO i}), \\
&H_r = \hbar \omega_r a^\dag a, \nonumber \\
&H_{DQD i} = \frac{1}{2}\qty(\epsilon_i \tau_{z i} + \Omega_i \tau_{x i}), \nonumber \\
&H_{z i} = \frac{1}{2} g^* \mu_B \vec{B_i} \vdot \vec{\sigma}_i = \frac{1}{2} \hbar \omega_{z i} \sigma_{z i}, \nonumber \\
&H_{SO i} = \frac{1}{4} g^* \mu_B \Delta\vec{B}_i\vdot\vec{\sigma}_i \tau_{z i} = \qty(g_{x i} \sigma_{x i} + g_{z i} \sigma_{z i}) \tau_{z i}, \nonumber \\
&H_{RO i} = e V_r \dyad{R}{R}_i = g_{AC i} \qty(a^\dag + a) \qty(1 - \tau_{z i}), \nonumber
\end{align}
where we have introduced the DQD detunings $\epsilon_i$ and tunneling constants $\Omega_i$ \cite{Hanson07}, the Zeeman splitting frequencies $\omega_{z i}$, the transverse and longitudinal spin-orbit coupling strengths $g_{x i}$ and $g_{z i}$, and the photon-orbit coupling strengths $g_{AC i}$ which are related to the resonator voltage $V_r$. We also introduce photonic creation and annihilation operators $a^\dag$ and $a$, the usual Pauli spin operators $\vec{\sigma_i} = \mqty(\sigma_{x i} & \sigma_{y i} & \sigma_{z i})$, as well as the orbital Pauli operators $\tau_{z_i} = \dyad{L}{L}_i - \dyad{R}{R}_i$ and $\tau_{x i} = \dyad{L}{R}_i + \dyad{R}{L}_i$.

At this point, we note that it is convenient to work in an orbital basis in which the $H_{DQD i}$ are diagonal, rather than the $\{\ket{L}_i,\ket{R}_i\}$ basis. We introduce the mixing angles $\theta_i = \arctan{\frac{\Omega_i}{\epsilon_i}}$ and move to the DQD eigenbasis $\ket{+}_i = \cos{\frac{\theta_i}{2}}\ket{L}_i + \sin{\frac{\theta_i}{2}}\ket{R}_i$ and $\ket{-}_i = \cos{\frac{\theta_i}{2}}\ket{R}_i - \sin{\frac{\theta_i}{2}}\ket{L}_i$. In this orbital basis, $H_{DQD_i} = \frac{1}{2} \hbar \omega_{a i} \tau_{z i}$ where $\omega_{a i} = \sqrt{\epsilon_i^2 + \Omega_i^2} / \hbar$. We label states of the complete system $\ket{\{s_1,s_2,...,s_N\},\{d_1,d_2,...,d_N\},n}$, where for each spin $s_i \in \{\uparrow,\downarrow\}$, each electron orbital state $d_i \in \{+,-\}$, and the photon number state is labeled $n \in \{0,1,2,...\}$.

We desire an effective Hamiltonian which describes the low-energy dynamics of the system, where we take low energy to mean the orbital degree of freedom is in its ground state (all $d_i = -$) and the cavity is unpopulated ($n = 0$). To derive such a Hamiltonian, we treat all $H_{SO i}$ and $H_{RO i}$ as perturbations, small relative to the remaining terms. We employ Schrieffer-Wolff transformations to eliminate to leading order all terms in $H_N$ which couple high- and low-energy states \cite{Schrieffer66}. This requires us to assume a separation in energy scales, e.g. $\omega_{a i} > \omega_r > \omega_{z i}$. We then project onto the low-energy subspace spanned by $\ket{\{s_1,s_2,...,s_N\},\{-,...,-\},0}$ to obtain the effective multi-spin Hamiltonian
\begin{align}
\label{eq:effHam}
H''_N &= \sum_{i = 1}^N \frac{1}{2} \hbar \omega''_{z i} \sigma_{z i} \\
	&- \sum_{i \neq j} \frac{\omega'_r}{\hbar} \qty(\frac{g'_{x j}}{\omega_{r}^{'2} - \omega_{z j}^{'2}}\sigma_{x j} + \frac{g'_{z j}}{\omega_{r}^{'2}}\sigma_{z j})\qty(g'_{x i} \sigma_{x i} + g'_{z i} \sigma_{z i}) \nonumber.
\end{align}
A full derivation of the effective Hamiltonian from Eq. (\ref{eq:baremultispin}) is given in the supplemental material \cite{Supplement}. Higher order terms can be safely ignored, assuming the $g_{x i}$, $g_{z i}$, and $g_{AC i}$ are sufficiently small relative to the differences in energy scales.

To demonstrate how Eq. (\ref{eq:effHam}) can be used to generate entangling operations, we focus on a system composed of $N=2$ DQDs with a purely transverse coupling (all $g_{z i} = 0$). The effective two-spin Hamiltonian becomes
\begin{align}
&H''_2 = \frac{1}{2} \hbar \omega''_{z 1} \sigma_{z 1} + \frac{1}{2} \hbar \omega''_{z 2} \sigma_{z 2} - J \sigma_{x 1} \sigma_{x 2}, \\
&J = \frac{\omega'_r g'_{x 1} g'_{x 2}}{\hbar} \qty(\frac{1}{\omega_r^{'2} - \omega_{z 1}^{'2}} + \frac{1}{\omega_r^{'2} - \omega_{z 2}^{'2}}). \nonumber
\end{align}

If we now transform into the rotating frame defined by $\omega''_{z 1}$ and $\omega''_{z 2}$ and drop counter-rotating terms, we are left with
\begin{equation}
\label{eq:xxHam}
\tilde{H} = -J\qty(\sigma_{- 1}\sigma_{+ 2}e^{i \Delta t} + \sigma_{- 2}\sigma_{+ 1} e^{-i \Delta t}),
\end{equation}
where $\sigma_{\pm i}$ are the spin raising/lowering operators, and we have defined the spin-spin detuning $\Delta = \omega''_{z 2} - \omega''_{z 1}$.

If the resonance condition $\Delta = 0$ is met, exponentiation yields the rotating frame time evolution operator
\begin{equation}
\label{eq:timeevolrot}
\tilde{U}(t) = \mqty(1 & 0 & 0 & 0 \\ 0 & \cos(\frac{J t}{\hbar}) & i \sin(\frac{J t}{\hbar}) & 0 \\ 0 & i \sin(\frac{J t}{\hbar}) & \cos(\frac{J t}{\hbar}) & 0 \\ 0 & 0 & 0 & 1).
\end{equation}
Time evolution under the Hamiltonian in Eq. (\ref{eq:xxHam}) generates the maximally entangling $iSWAP$ gate when $J t / \hbar = \pi / 2$, or the perfectly entangling $\sqrt{iSWAP}$ gate in half the time \cite{McKay16, Rezakhani04, Srinivasa16}. When the spins are detuned, however ($\abs{\hbar\Delta} >> \abs{J}$), the terms in Eq. (\ref{eq:xxHam}) oscillate so quickly that they become negligible, and spin evolution in the rotating frame becomes trivial.

Crucially, both $\Delta$ and $J$ depend on the $\epsilon_i$ and $\Omega_i$. This means both dressed spin energy splittings and effective spin-spin couplings are electrically controllable. Though Eq. (\ref{eq:effHam}) was derived assuming no time dependence for $H_N$, it remains a useful approximation in cases where the $\epsilon_i$ are time-dependent, so long as the time dependence is sufficiently small as to avoid Landau-Zener transitions to the higher-energy subspace \cite{Zener32}. As DQD gate voltages can be manipulated on very short time scales \cite{Petta05, Hanson07}, this suggests the possibility of fast, on-demand entanglement generation by simply adjusting the $\epsilon_i$, moving distant spins into and out of resonance with one another.

In order to test the validity of the effective Hamiltonian in Eq. (\ref{eq:effHam}), we compare it to numerical simulations that use the original Hamiltonian in Eq. (\ref{eq:baremultispin}). Our simulations include leakage to higher-energy states. We truncate to ten photonic states, which is more than sufficient to accurately obtain the system's low-energy dynamics. We focus on two DQD systems with physically realistic parameters taken from Ref. \cite{Mi18}.

\begin{figure}
\includegraphics[width=0.9\textwidth]{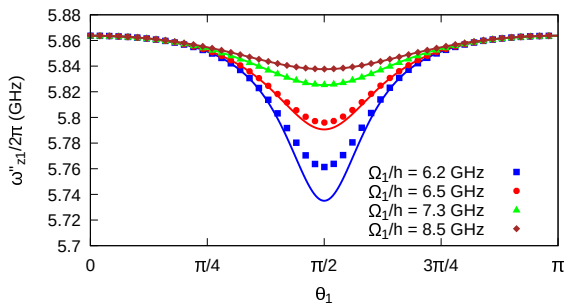}
\caption{A plot of the analytically (solid lines) and numerically (markers) calculated spin splitting $\omega''_{z1}$ as $\epsilon_1$ is adjusted for various fixed values of $\Omega_1$. Here we have set $\omega_r/2\pi = \SI{6}{\giga\hertz}$, $\omega_{z1}/2\pi = \omega_{z2}/2\pi = \SI{5.85}{\giga\hertz}$, $\Omega_2 / h = \SI{7.5}{\giga\hertz}$, $g_{x1}/h = g_{x2}/h = \SI{200}{\mega\hertz}$, $g_{AC1} / h = g_{AC2}/h = \SI{40}{\mega\hertz}$, and $\theta_2 = \pi/2$.}
\label{fig:energies}
\end{figure}

To investigate the low-energy dressed spin splittings, we diagonalize the multi-DQD Hamiltonian in Eq. (\ref{eq:baremultispin}) and look at eigenvalues whose eigenstates strongly overlap with the unperturbed low-energy subspace states ($d_i = -$, $n = 0$). So long as our energy scale separation assumptions hold, each unperturbed low-energy state will have a large overlap with exactly one eigenstate of the Hamiltonian \cite{Bravyi11}. Fig. \ref{fig:energies} compares numerically calculated spin energy splittings against analytic predictions as the DQD detuning $\epsilon_1$ is varied with other parameters held constant. For large $\Omega_1$ relative to $\omega_r$ and $\omega_{z1}$, agreement between numerical results and analytic predictions is very good. As $\Omega_1$ is decreased, however, our energy scale assumptions become less sound. The approximation begins to break down, and neglected terms in Eq. (\ref{eq:effHam}) become significant, leading to a relatively large error near $\theta_1 = \pi/2$.

\begin{figure}
\includegraphics[width=0.9\textwidth]{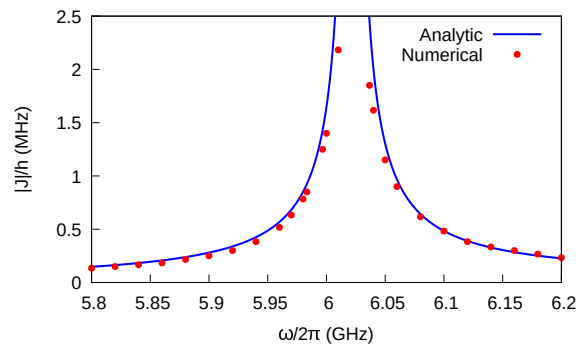}
\caption{Plot of effective coupling strength $J$ vs the spin splittings $\omega_{z1} = \omega_{z2} = \omega$. Here we have set $\omega_r/2\pi = \SI{6}{\giga\hertz}$, $\Omega_1/h = \Omega_2/h = \SI{7.5}{\giga\hertz}$, $g_{x 1}/h = g_{x 2}/h = \SI{200}{\mega\hertz}$, $g_{AC1}/h = g_{AC2}/h = \SI{40}{\mega\hertz}$, and $\theta_1 = \theta_2 = \pi/2$. Near $\omega/2\pi = \SI{6.02}{\giga\hertz}$, the spins are brought into resonance with the cavity, and our approximations break down.}
\label{fig:Js}
\end{figure}

The effective coupling $J$ is most easily extracted by numerically computing the $\ket{\uparrow \downarrow} \leftrightarrow \ket{\downarrow \uparrow}$ transition probability once the spins have been brought into resonance. From Eq. (\ref{eq:timeevolrot}), this transition probability is clearly proportional to $\sin[2](J t / \hbar)$. By looking at the frequency of transition probability oscillations, therefore, we obtain an estimate of $J$. These numerically-determined effective couplings are plotted against analytic predictions in Fig. \ref{fig:Js}. Again, agreement is quite good when our assumptions about energy separation hold. 

We now turn to simulations of on-demand, electrically generated entangling gates. As stated above, entanglement generation can essentially be turned on or off by bringing the spins into or out of resonance with one another. We use the effective Hamiltonian of Eq. \ref{eq:effHam} to choose parameters such that spin-spin resonance can be achieved by changing gate voltages alone. We aim to achieve resonance close to $\theta_1 = \theta_2 = \pi/2$ to maximize $J$. When the spins are detuned from one another, time evolution in the rotating frame becomes trivial. It is convenient, then, to work in this rotating frame rather than the lab frame where spin evolution is always nontrivial.

\begin{figure}
\includegraphics[width=0.95\linewidth]{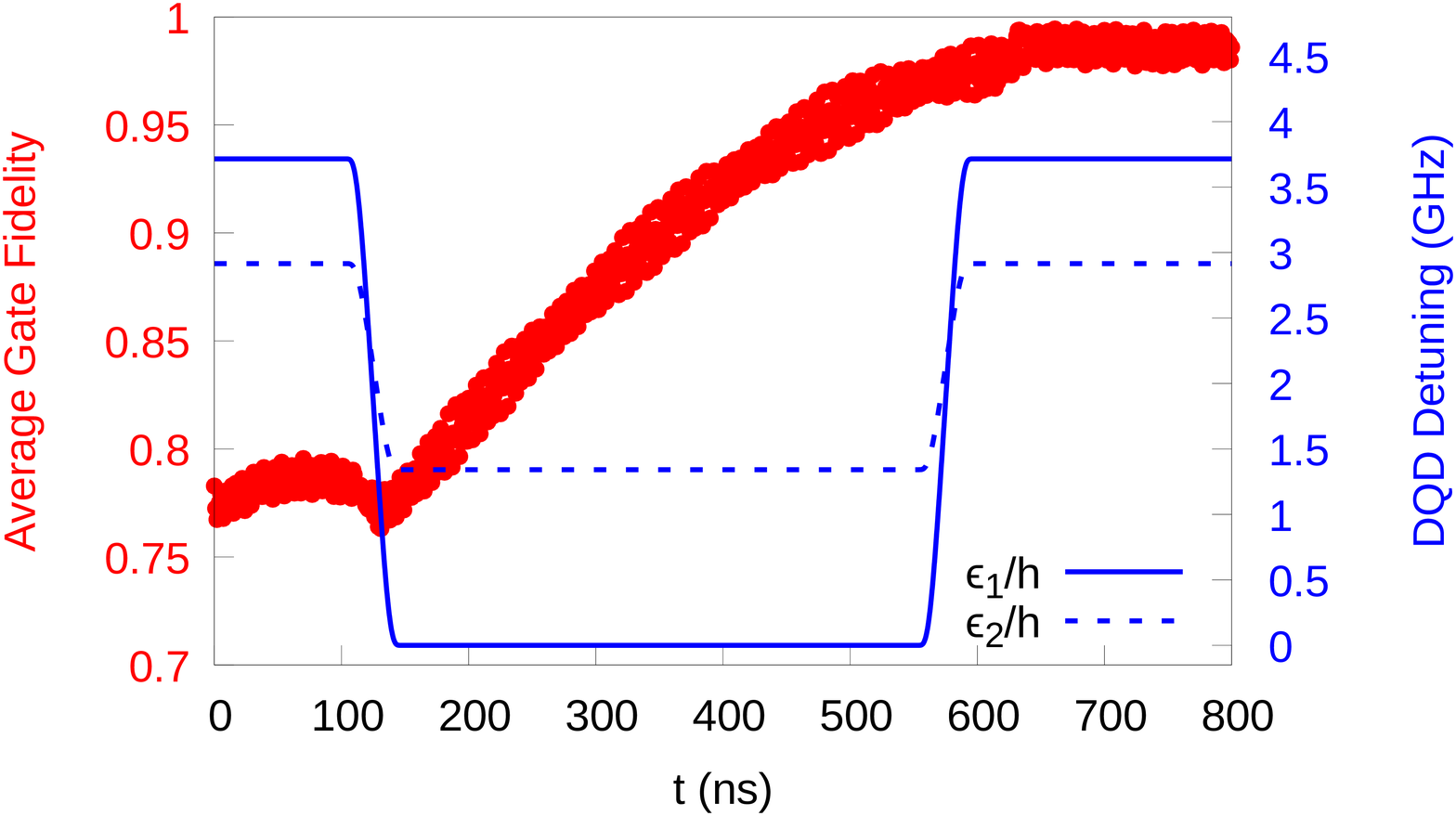}
\caption{For this simulation, we set $\omega_r/2\pi = \SI{6}{\giga\hertz}$, $\omega_{z1}/2\pi = \omega_{z2}/2\pi = \SI{5.86}{\giga\hertz}$, $\Omega_1 / h = \SI{7.3}{\giga\hertz}$, $\Omega_2 / h = \SI{7.5}{\giga\hertz}$, $g_{AC1} / h = \SI{45}{\mega\hertz}$, $g_{AC2} / h = \SI{40}{\mega\hertz}$, $g_{x 1} = \SI{200}{\mega\hertz}$, and $g_{x 2} = \SI{230}{\mega\hertz}$. Here we plot the $\epsilon_i$ used to generate an entangling gate, along with the average gate fidelity of the spins' time evolution relative to a $\sqrt{iSWAP}$, maximized at each time step over local operations. Beginning at $t = \SI{100}{\nano\second}$, we smoothly decrease both detunings to bring the spins into resonance. Then, starting at $t = \SI{550}{\nano\second}$, we smoothly transition back. Total leakage does not exceed $3\%$ after $t = \SI{600}{\nano\second}$.}
\label{fig:fidelity}
\end{figure}

We numerically solve the Schr\"{o}dinger equation for the Hamiltonian in Eq. (\ref{eq:baremultispin}) where the $\epsilon_i$ are now time-dependent. At each time step, we calculate the time evolution of low-energy spin states, compute corresponding density operators, and then trace out the resonator and orbital degrees of freedom. Our gate fidelities are computed according to the definition given in \cite{Nielsen02}: $\bar{F}(\mathcal{E}, U) = \frac{1}{5} + \frac{1}{80}\sum_{j,k=\mathbbm{1},x,y,z} \tr(U \sigma_{j1}\sigma_{k2} U^\dag \mathcal{E}(\sigma_{j1}\sigma_{k2}))$ where $U$ is our target gate and $\mathcal{E}(\rho)$ is the quantum process representing the spin time evolution described above. We choose asymmetric parameter values for the DQD systems (see Fig. \ref{fig:fidelity} caption) to illustrate the robustness of the electrical control provided by the architecture.

We start our simulation with the spins well-detuned from one another, $\theta_1 = 1.1$ and $\theta_2 = 1.2$. After $\SI{100}{\nano\second}$, we vary the $\epsilon_i$ to bring the spins into resonance near $\theta_1 = \pi/2$ and $\theta_2 = 1.3983$. To avoid Landau-Zener transitions, we must avoid modulating system parameters too rapidly. We transition between non-resonant and resonant configurations slowly, over $\SI{50}{\nano\second}$, interpolating with the smooth transition function $g(x) = \frac{e^{-1/x}}{e^{-1/x} + e^{-1/(1-x)}}.$

We maintain resonance for $\SI{400}{\nano\second}$ before transitioning back over $\SI{50}{\nano\second}$ to the original detuned parameters. The full control sequence is shown in Fig. \ref{fig:fidelity}, along with the fidelity as a function of time. After the $\SI{500}{\nano\second}$ pulse sequence, we have generated a gate locally equivalent to a $\sqrt{iSWAP}$ with between $97.5\%$ and $99.5\%$ fidelity, with leakage to higher-energy states oscillating under $3\%$. We choose a local equivalent rather than the $\sqrt{iSWAP}$ itself because the change in spin splitting associated with tuning to resonance generates local $z$-rotations in the rotating frame. This can be used to generate phase gates by momentarily changing DQD detunings without bringing the spins into resonance. In principle we could use this mechanism to eliminate local phases and generate the $\sqrt{iSWAP}$ itself, rather than a locally equivalent gate.

So far we have discussed only noiseless systems, but fidelities achievable in real solid-state spin systems are limited by charge noise, phonon interactions, and cavity loss \cite{Mi18}. The effects of low-frequency, quasistatic charge noise can be modeled by making the substitutions $\Omega_i \rightarrow \Omega_i + \delta\Omega_i$ and $\epsilon_i \rightarrow \epsilon_i + \delta\epsilon_i$, where the $\delta\Omega_i$ and $\delta\epsilon_i$ are Gaussian-distributed random variables with standard deviation $\sigma_\epsilon$. For the purposes of two-qubit gate generation, the primary effect of this quasistatic noise is to change the spin-spin detuning $\Delta$ and the effective spin-spin coupling $J$.

\begin{figure}
\includegraphics[width=0.9\linewidth]{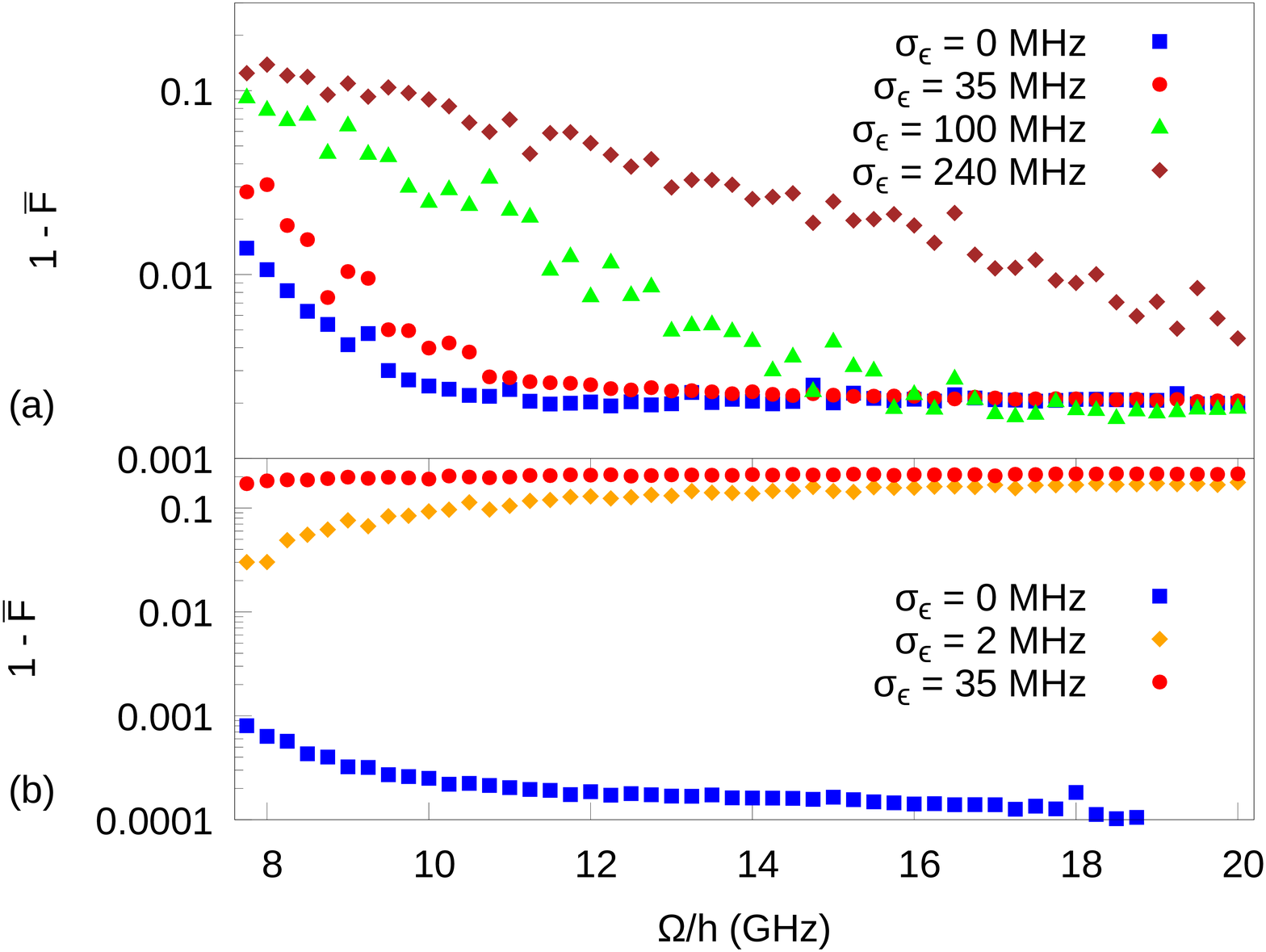}
\caption{Average infidelity of $\sqrt{iSWAP}$ gates for various values of $\sigma_\epsilon$. For each noise instance, at each time step we compute the average gate fidelity maximized over local operations. We then average over noise instances the minimum of the gate infidelity as a function of time. Here, we have set $\omega_r/2\pi = \SI{6}{\giga\hertz}$, $\Omega_1 = \Omega_2 = \Omega$, $g_{AC1}/h = g_{AC2}/h = \SI{40}{\mega\hertz}$, and $\theta_1 = \theta_2 = \pi/2$. (a) A plot of the infidelity for the spin qubit, where we have set $\omega_{z 1} = \omega_{z 2} = \omega$ and $g_{x 1}/h = g_{x 2}/h = \SI{200}{\mega\hertz}$. To ensure that spins remain dispersively coupled to the resonator, at each $\Omega$, we choose $\omega$ such that $\omega'_{r} - \omega'_{z i} = 30 g'_{x i}$ (definitions given in supplemental material \cite{Supplement}). Average gate times increase from $\SI{830}{\nano\second}$ at $\Omega/h = \SI{8}{\giga\hertz}$ to $\SI{4.3}{\micro\second}$ at $\Omega/h = \SI{20}{\giga\hertz}$ (b) A plot of the infidelity for the charge qubit, where we have set $g_{x 1} = g_{x 2} = 0$. Average gate times for the charge qubit are shorter than for the spin qubit, ranging from $\SI{180}{\nano\second}$ at $\Omega/h = \SI{8}{\giga\hertz}$ to $\SI{2.4}{\micro\second}$ at $\Omega/h = \SI{20}{\giga\hertz}$.}
\label{fig:chargenoise}
\end{figure}

We plot average $\sqrt{iSWAP}$ gate fidelities achievable for various values of $\sigma_\epsilon$ in Fig. \ref{fig:chargenoise}(a). For comparison, we also plot $\sqrt{iSWAP}$ gate fidelities achievable using the orbital degrees of freedom as our qubits in Fig. \ref{fig:chargenoise}(b). Previous experimental work measured charge noise levels at around $\SI{35}{\mega\hertz}$ \cite{Mi18} and as low as $\SI{2.6}{\mega\hertz}$ \cite{Mi17}. We note that at these levels of quasistatic charge noise, the spin qubits' gate performance is not significantly worse than the noiseless case, and at all nonzero $\sigma_\epsilon$, spin qubits achieve higher fidelities than charge qubits for a given set of orbital parameters.

For smaller $\Omega_i$, closer to $\omega_r$ and the $\omega_{z i}$, the average gate fidelity for spin qubits is more sensitive to quasistatic charge noise fluctuations, as $\abs{\pdv{\Delta}{\omega_{a i}}}$ and $\abs{\pdv{J}{\omega_{a i}}}$ become larger. This regime is also where we expect high-frequency charge noise and phonon-induced dephasing to have the largest effect \cite{Mi18}. As the $\Omega_i$ are increased, the spin qubits become less sensitive to changes in orbital parameters, and the fidelities converge towards the noiseless value. The remaining infidelity at $\sigma_\epsilon = 0$ is due to leakage. For smaller values of $\Omega_i$, where $\abs{\frac{g_{x i}}{\omega_{a i} - \omega_{z i}}}$ is relatively large, leakage to the excited orbital subspace dominates, whereas for larger values of $\Omega_i$, leakage to excited cavity states becomes the dominant source of infidelity.

Although increasing the $\Omega_i$ has the effect of decreasing orbital leakage and charge noise sensitivity, it also decreases $J$, increasing two-qubit gate times. As gate time increases, cavity loss will become a greater source of infidelity, as qubits must remain strongly coupled to the cavity for longer in order to generate entanglement. The competing nature of these sources of infidelity should give rise to an optimal $\Omega$ at which gate performance is maximal. This is investigated further in \cite{Benito19}.

Our two-qubit gate time is less than spin coherence lifetimes reported in Ref. \cite{Mi18}, suggesting the possibility of experimental realization in the near future. Additionally, the full effective Hamiltonian in Eq. (\ref{eq:effHam}) indicates the possibility of generating other multi-qubit gates, e.g. by making use of longitudinal couplings, by coupling more DQD systems to the same resonator, or by using more complicated driving fields \cite{Srinivasa16}. These entangling gates, along with local phase gates and already experimentally realized EDSR rotations, would compose a set of universal quantum gates, all of which could be generated purely via electrical manipulation. These gate generation mechanisms and our ability to utilize long-range effective spin-spin interactions greatly enhance prospects of using solid-state DQD electronic spins for quantum information processing.

\textit{Acknowledgments:} this work is supported by the Army Research Office (W911NF-17-0287).

\bibliography{refs}{}

\begin{thebibliography}{35}%
\makeatletter
\providecommand \@ifxundefined [1]{%
 \@ifx{#1\undefined}
}%
\providecommand \@ifnum [1]{%
 \ifnum #1\expandafter \@firstoftwo
 \else \expandafter \@secondoftwo
 \fi
}%
\providecommand \@ifx [1]{%
 \ifx #1\expandafter \@firstoftwo
 \else \expandafter \@secondoftwo
 \fi
}%
\providecommand \natexlab [1]{#1}%
\providecommand \enquote  [1]{``#1''}%
\providecommand \bibnamefont  [1]{#1}%
\providecommand \bibfnamefont [1]{#1}%
\providecommand \citenamefont [1]{#1}%
\providecommand \href@noop [0]{\@secondoftwo}%
\providecommand \href [0]{\begingroup \@sanitize@url \@href}%
\providecommand \@href[1]{\@@startlink{#1}\@@href}%
\providecommand \@@href[1]{\endgroup#1\@@endlink}%
\providecommand \@sanitize@url [0]{\catcode `\\12\catcode `\$12\catcode
  `\&12\catcode `\#12\catcode `\^12\catcode `\_12\catcode `\%12\relax}%
\providecommand \@@startlink[1]{}%
\providecommand \@@endlink[0]{}%
\providecommand \url  [0]{\begingroup\@sanitize@url \@url }%
\providecommand \@url [1]{\endgroup\@href {#1}{\urlprefix }}%
\providecommand \urlprefix  [0]{URL }%
\providecommand \Eprint [0]{\href }%
\providecommand \doibase [0]{http://dx.doi.org/}%
\providecommand \selectlanguage [0]{\@gobble}%
\providecommand \bibinfo  [0]{\@secondoftwo}%
\providecommand \bibfield  [0]{\@secondoftwo}%
\providecommand \translation [1]{[#1]}%
\providecommand \BibitemOpen [0]{}%
\providecommand \bibitemStop [0]{}%
\providecommand \bibitemNoStop [0]{.\EOS\space}%
\providecommand \EOS [0]{\spacefactor3000\relax}%
\providecommand \BibitemShut  [1]{\csname bibitem#1\endcsname}%
\let\auto@bib@innerbib\@empty
\bibitem [{\citenamefont {Kawakami}\ \emph {et~al.}(2016)\citenamefont
  {Kawakami}, \citenamefont {Jullien}, \citenamefont {Scarlino}, \citenamefont
  {Ward}, \citenamefont {Savage}, \citenamefont {Lagally}, \citenamefont
  {Dobrovitski}, \citenamefont {Friesen}, \citenamefont {Coppersmith},
  \citenamefont {Eriksson},\ and\ \citenamefont {Vandersypen}}]{Kawakami16}%
  \BibitemOpen
  \bibfield  {author} {\bibinfo {author} {\bibfnamefont {E.}~\bibnamefont
  {Kawakami}}, \bibinfo {author} {\bibfnamefont {T.}~\bibnamefont {Jullien}},
  \bibinfo {author} {\bibfnamefont {P.}~\bibnamefont {Scarlino}}, \bibinfo
  {author} {\bibfnamefont {D.~R.}\ \bibnamefont {Ward}}, \bibinfo {author}
  {\bibfnamefont {D.~E.}\ \bibnamefont {Savage}}, \bibinfo {author}
  {\bibfnamefont {M.~G.}\ \bibnamefont {Lagally}}, \bibinfo {author}
  {\bibfnamefont {V.~V.}\ \bibnamefont {Dobrovitski}}, \bibinfo {author}
  {\bibfnamefont {M.}~\bibnamefont {Friesen}}, \bibinfo {author} {\bibfnamefont
  {S.~N.}\ \bibnamefont {Coppersmith}}, \bibinfo {author} {\bibfnamefont
  {M.~A.}\ \bibnamefont {Eriksson}}, \ and\ \bibinfo {author} {\bibfnamefont
  {L.~M.~K.}\ \bibnamefont {Vandersypen}},\ }\href {\doibase
  10.1073/pnas.1603251113} {\bibfield  {journal} {\bibinfo  {journal}
  {Proceedings of the National Academy of Sciences}\ }\textbf {\bibinfo
  {volume} {113}},\ \bibinfo {pages} {11738} (\bibinfo {year}
  {2016})}\BibitemShut {NoStop}%
\bibitem [{\citenamefont {Veldhorst}\ \emph {et~al.}(2014)\citenamefont
  {Veldhorst}, \citenamefont {Hwang}, \citenamefont {Yang}, \citenamefont
  {Leenstra}, \citenamefont {de~Ronde}, \citenamefont {Dehollain},
  \citenamefont {Muhonen}, \citenamefont {Hudson}, \citenamefont {Itoh},
  \citenamefont {Morello},\ and\ \citenamefont {Dzurak}}]{Veldhorst14}%
  \BibitemOpen
  \bibfield  {author} {\bibinfo {author} {\bibfnamefont {M.}~\bibnamefont
  {Veldhorst}}, \bibinfo {author} {\bibfnamefont {J.~C.~C.}\ \bibnamefont
  {Hwang}}, \bibinfo {author} {\bibfnamefont {C.~H.}\ \bibnamefont {Yang}},
  \bibinfo {author} {\bibfnamefont {A.~W.}\ \bibnamefont {Leenstra}}, \bibinfo
  {author} {\bibfnamefont {B.}~\bibnamefont {de~Ronde}}, \bibinfo {author}
  {\bibfnamefont {J.~P.}\ \bibnamefont {Dehollain}}, \bibinfo {author}
  {\bibfnamefont {J.~T.}\ \bibnamefont {Muhonen}}, \bibinfo {author}
  {\bibfnamefont {F.~E.}\ \bibnamefont {Hudson}}, \bibinfo {author}
  {\bibfnamefont {K.~M.}\ \bibnamefont {Itoh}}, \bibinfo {author}
  {\bibfnamefont {A.}~\bibnamefont {Morello}}, \ and\ \bibinfo {author}
  {\bibfnamefont {A.~S.}\ \bibnamefont {Dzurak}},\ }\href {\doibase
  10.1038/nnano.2014.216} {\bibfield  {journal} {\bibinfo  {journal} {Nature
  Nanotechnology}\ }\textbf {\bibinfo {volume} {9}},\ \bibinfo {pages} {981}
  (\bibinfo {year} {2014})}\BibitemShut {NoStop}%
\bibitem [{\citenamefont {Zwanenburg}\ \emph {et~al.}(2013)\citenamefont
  {Zwanenburg}, \citenamefont {Dzurak}, \citenamefont {Morello}, \citenamefont
  {Simmons}, \citenamefont {Hollenberg}, \citenamefont {Klimeck}, \citenamefont
  {Rogge}, \citenamefont {Coppersmith},\ and\ \citenamefont
  {Eriksson}}]{Zwanenburg13}%
  \BibitemOpen
  \bibfield  {author} {\bibinfo {author} {\bibfnamefont {F.~A.}\ \bibnamefont
  {Zwanenburg}}, \bibinfo {author} {\bibfnamefont {A.~S.}\ \bibnamefont
  {Dzurak}}, \bibinfo {author} {\bibfnamefont {A.}~\bibnamefont {Morello}},
  \bibinfo {author} {\bibfnamefont {M.~Y.}\ \bibnamefont {Simmons}}, \bibinfo
  {author} {\bibfnamefont {L.~C.~L.}\ \bibnamefont {Hollenberg}}, \bibinfo
  {author} {\bibfnamefont {G.}~\bibnamefont {Klimeck}}, \bibinfo {author}
  {\bibfnamefont {S.}~\bibnamefont {Rogge}}, \bibinfo {author} {\bibfnamefont
  {S.~N.}\ \bibnamefont {Coppersmith}}, \ and\ \bibinfo {author} {\bibfnamefont
  {M.~A.}\ \bibnamefont {Eriksson}},\ }\href {\doibase
  10.1103/RevModPhys.85.961} {\bibfield  {journal} {\bibinfo  {journal} {Rev.
  Mod. Phys.}\ }\textbf {\bibinfo {volume} {85}},\ \bibinfo {pages} {961}
  (\bibinfo {year} {2013})}\BibitemShut {NoStop}%
\bibitem [{\citenamefont {Loss}\ and\ \citenamefont
  {DiVincenzo}(1998)}]{Loss98}%
  \BibitemOpen
  \bibfield  {author} {\bibinfo {author} {\bibfnamefont {D.}~\bibnamefont
  {Loss}}\ and\ \bibinfo {author} {\bibfnamefont {D.~P.}\ \bibnamefont
  {DiVincenzo}},\ }\href {\doibase 10.1103/PhysRevA.57.120} {\bibfield
  {journal} {\bibinfo  {journal} {Phys. Rev. A}\ }\textbf {\bibinfo {volume}
  {57}},\ \bibinfo {pages} {120} (\bibinfo {year} {1998})}\BibitemShut
  {NoStop}%
\bibitem [{\citenamefont {Dehollain}\ \emph {et~al.}(2016)\citenamefont
  {Dehollain}, \citenamefont {Simmons}, \citenamefont {Muhonen}, \citenamefont
  {Kalra}, \citenamefont {Laucht}, \citenamefont {Hudson}, \citenamefont
  {Itoh}, \citenamefont {Jamieson}, \citenamefont {McCallum}, \citenamefont
  {Dzurak},\ and\ \citenamefont {Morello}}]{Dehollain16}%
  \BibitemOpen
  \bibfield  {author} {\bibinfo {author} {\bibfnamefont {J.~P.}\ \bibnamefont
  {Dehollain}}, \bibinfo {author} {\bibfnamefont {S.}~\bibnamefont {Simmons}},
  \bibinfo {author} {\bibfnamefont {J.~T.}\ \bibnamefont {Muhonen}}, \bibinfo
  {author} {\bibfnamefont {R.}~\bibnamefont {Kalra}}, \bibinfo {author}
  {\bibfnamefont {A.}~\bibnamefont {Laucht}}, \bibinfo {author} {\bibfnamefont
  {F.}~\bibnamefont {Hudson}}, \bibinfo {author} {\bibfnamefont
  {K.}~\bibnamefont {Itoh}}, \bibinfo {author} {\bibfnamefont {D.~N.}\
  \bibnamefont {Jamieson}}, \bibinfo {author} {\bibfnamefont {J.~C.}\
  \bibnamefont {McCallum}}, \bibinfo {author} {\bibfnamefont {A.~S.}\
  \bibnamefont {Dzurak}}, \ and\ \bibinfo {author} {\bibfnamefont
  {A.}~\bibnamefont {Morello}},\ }\href {\doibase 10.1038/NNANO.2015.262}
  {\bibfield  {journal} {\bibinfo  {journal} {Nature Nanotechnology}\ }\textbf
  {\bibinfo {volume} {11}},\ \bibinfo {pages} {242} (\bibinfo {year}
  {2016})}\BibitemShut {NoStop}%
\bibitem [{\citenamefont {Ladd}\ \emph {et~al.}(2010)\citenamefont {Ladd},
  \citenamefont {Jelezko}, \citenamefont {Laflamme}, \citenamefont {Nakamura},
  \citenamefont {Monroe},\ and\ \citenamefont {O'Brien}}]{Ladd10}%
  \BibitemOpen
  \bibfield  {author} {\bibinfo {author} {\bibfnamefont {T.~D.}\ \bibnamefont
  {Ladd}}, \bibinfo {author} {\bibfnamefont {F.}~\bibnamefont {Jelezko}},
  \bibinfo {author} {\bibfnamefont {R.}~\bibnamefont {Laflamme}}, \bibinfo
  {author} {\bibfnamefont {Y.}~\bibnamefont {Nakamura}}, \bibinfo {author}
  {\bibfnamefont {C.}~\bibnamefont {Monroe}}, \ and\ \bibinfo {author}
  {\bibfnamefont {J.~L.}\ \bibnamefont {O'Brien}},\ }\href {\doibase
  10.1038/nature08812} {\bibfield  {journal} {\bibinfo  {journal} {Nature}\
  }\textbf {\bibinfo {volume} {464}},\ \bibinfo {pages} {45} (\bibinfo {year}
  {2010})}\BibitemShut {NoStop}%
\bibitem [{\citenamefont {Petersson}\ \emph {et~al.}(2012)\citenamefont
  {Petersson}, \citenamefont {McFaul}, \citenamefont {Schroer}, \citenamefont
  {Jung}, \citenamefont {Taylor}, \citenamefont {Houck},\ and\ \citenamefont
  {Petta}}]{Petersson12}%
  \BibitemOpen
  \bibfield  {author} {\bibinfo {author} {\bibfnamefont {K.~D.}\ \bibnamefont
  {Petersson}}, \bibinfo {author} {\bibfnamefont {L.~W.}\ \bibnamefont
  {McFaul}}, \bibinfo {author} {\bibfnamefont {M.~D.}\ \bibnamefont {Schroer}},
  \bibinfo {author} {\bibfnamefont {M.}~\bibnamefont {Jung}}, \bibinfo {author}
  {\bibfnamefont {J.~M.}\ \bibnamefont {Taylor}}, \bibinfo {author}
  {\bibfnamefont {A.~A.}\ \bibnamefont {Houck}}, \ and\ \bibinfo {author}
  {\bibfnamefont {J.~R.}\ \bibnamefont {Petta}},\ }\href {\doibase
  10.1038/nature11559} {\bibfield  {journal} {\bibinfo  {journal} {Nature}\
  }\textbf {\bibinfo {volume} {490}},\ \bibinfo {pages} {380} (\bibinfo {year}
  {2012})}\BibitemShut {NoStop}%
\bibitem [{\citenamefont {Blais}\ \emph {et~al.}(2007)\citenamefont {Blais},
  \citenamefont {Gambetta}, \citenamefont {Wallraff}, \citenamefont {Schuster},
  \citenamefont {Girvin}, \citenamefont {Devoret},\ and\ \citenamefont
  {Schoelkopf}}]{Blais07}%
  \BibitemOpen
  \bibfield  {author} {\bibinfo {author} {\bibfnamefont {A.}~\bibnamefont
  {Blais}}, \bibinfo {author} {\bibfnamefont {J.}~\bibnamefont {Gambetta}},
  \bibinfo {author} {\bibfnamefont {A.}~\bibnamefont {Wallraff}}, \bibinfo
  {author} {\bibfnamefont {D.~I.}\ \bibnamefont {Schuster}}, \bibinfo {author}
  {\bibfnamefont {S.~M.}\ \bibnamefont {Girvin}}, \bibinfo {author}
  {\bibfnamefont {M.~H.}\ \bibnamefont {Devoret}}, \ and\ \bibinfo {author}
  {\bibfnamefont {R.~J.}\ \bibnamefont {Schoelkopf}},\ }\href {\doibase
  10.1103/PhysRevA.75.032329} {\bibfield  {journal} {\bibinfo  {journal} {Phys.
  Rev. A}\ }\textbf {\bibinfo {volume} {75}},\ \bibinfo {pages} {032329}
  (\bibinfo {year} {2007})}\BibitemShut {NoStop}%
\bibitem [{\citenamefont {Schoelkopf}\ and\ \citenamefont
  {Girvin}(2008)}]{Schoelkopf08}%
  \BibitemOpen
  \bibfield  {author} {\bibinfo {author} {\bibfnamefont {R.~J.}\ \bibnamefont
  {Schoelkopf}}\ and\ \bibinfo {author} {\bibfnamefont {S.~M.}\ \bibnamefont
  {Girvin}},\ }\href {\doibase 10.1038/451664a} {\bibfield  {journal} {\bibinfo
   {journal} {Nature}\ }\textbf {\bibinfo {volume} {451}},\ \bibinfo {pages}
  {664} (\bibinfo {year} {2008})}\BibitemShut {NoStop}%
\bibitem [{\citenamefont {Ams\"{u}ss}\ \emph {et~al.}(2011)\citenamefont
  {Ams\"{u}ss}, \citenamefont {Koller}, \citenamefont {N\"{o}bauer},
  \citenamefont {Putz}, \citenamefont {Rotter}, \citenamefont {Sandner},
  \citenamefont {Schneider}, \citenamefont {Schramb\"{o}ck}, \citenamefont
  {Steinhauser}, \citenamefont {Ritsch}, \citenamefont {Schmiedmayer},\ and\
  \citenamefont {Majer}}]{Amsuss11}%
  \BibitemOpen
  \bibfield  {author} {\bibinfo {author} {\bibfnamefont {R.}~\bibnamefont
  {Ams\"{u}ss}}, \bibinfo {author} {\bibfnamefont {C.}~\bibnamefont {Koller}},
  \bibinfo {author} {\bibfnamefont {T.}~\bibnamefont {N\"{o}bauer}}, \bibinfo
  {author} {\bibfnamefont {S.}~\bibnamefont {Putz}}, \bibinfo {author}
  {\bibfnamefont {S.}~\bibnamefont {Rotter}}, \bibinfo {author} {\bibfnamefont
  {K.}~\bibnamefont {Sandner}}, \bibinfo {author} {\bibfnamefont
  {S.}~\bibnamefont {Schneider}}, \bibinfo {author} {\bibfnamefont
  {M.}~\bibnamefont {Schramb\"{o}ck}}, \bibinfo {author} {\bibfnamefont
  {G.}~\bibnamefont {Steinhauser}}, \bibinfo {author} {\bibfnamefont
  {H.}~\bibnamefont {Ritsch}}, \bibinfo {author} {\bibfnamefont
  {J.}~\bibnamefont {Schmiedmayer}}, \ and\ \bibinfo {author} {\bibfnamefont
  {J.}~\bibnamefont {Majer}},\ }\href {\doibase 10.1103/PhysRevLett.107.060502}
  {\bibfield  {journal} {\bibinfo  {journal} {Phys. Rev. Lett.}\ }\textbf
  {\bibinfo {volume} {107}},\ \bibinfo {pages} {060502} (\bibinfo {year}
  {2011})}\BibitemShut {NoStop}%
\bibitem [{\citenamefont {Schuster}\ \emph {et~al.}(2010)\citenamefont
  {Schuster}, \citenamefont {Sears}, \citenamefont {Ginossar}, \citenamefont
  {DiCarlo}, \citenamefont {Frunzio}, \citenamefont {Morton}, \citenamefont
  {Wu}, \citenamefont {Briggs}, \citenamefont {Buckley}, \citenamefont
  {Awschalom},\ and\ \citenamefont {Schoelkopf}}]{Schuster10}%
  \BibitemOpen
  \bibfield  {author} {\bibinfo {author} {\bibfnamefont {D.~I.}\ \bibnamefont
  {Schuster}}, \bibinfo {author} {\bibfnamefont {A.~P.}\ \bibnamefont {Sears}},
  \bibinfo {author} {\bibfnamefont {E.}~\bibnamefont {Ginossar}}, \bibinfo
  {author} {\bibfnamefont {L.}~\bibnamefont {DiCarlo}}, \bibinfo {author}
  {\bibfnamefont {L.}~\bibnamefont {Frunzio}}, \bibinfo {author} {\bibfnamefont
  {J.~J.~L.}\ \bibnamefont {Morton}}, \bibinfo {author} {\bibfnamefont
  {H.}~\bibnamefont {Wu}}, \bibinfo {author} {\bibfnamefont {G.~A.~D.}\
  \bibnamefont {Briggs}}, \bibinfo {author} {\bibfnamefont {B.~B.}\
  \bibnamefont {Buckley}}, \bibinfo {author} {\bibfnamefont {D.~D.}\
  \bibnamefont {Awschalom}}, \ and\ \bibinfo {author} {\bibfnamefont {R.~J.}\
  \bibnamefont {Schoelkopf}},\ }\href {\doibase 10.1103/PhysRevLett.105.140501}
  {\bibfield  {journal} {\bibinfo  {journal} {Phys. Rev. Lett.}\ }\textbf
  {\bibinfo {volume} {105}},\ \bibinfo {pages} {140501} (\bibinfo {year}
  {2010})}\BibitemShut {NoStop}%
\bibitem [{\citenamefont {Viennot}\ \emph {et~al.}(2015)\citenamefont
  {Viennot}, \citenamefont {Dartiailh}, \citenamefont {Cottet},\ and\
  \citenamefont {Kontos}}]{Viennot15}%
  \BibitemOpen
  \bibfield  {author} {\bibinfo {author} {\bibfnamefont {J.~J.}\ \bibnamefont
  {Viennot}}, \bibinfo {author} {\bibfnamefont {M.~C.}\ \bibnamefont
  {Dartiailh}}, \bibinfo {author} {\bibfnamefont {A.}~\bibnamefont {Cottet}}, \
  and\ \bibinfo {author} {\bibfnamefont {T.}~\bibnamefont {Kontos}},\ }\href
  {\doibase 10.1126/science.aaa3786} {\bibfield  {journal} {\bibinfo  {journal}
  {Science}\ }\textbf {\bibinfo {volume} {349}},\ \bibinfo {pages} {408}
  (\bibinfo {year} {2015})},\ \Eprint
  {http://arxiv.org/abs/https://science.sciencemag.org/content/349/6246/408.full.pdf}
  {https://science.sciencemag.org/content/349/6246/408.full.pdf} \BibitemShut
  {NoStop}%
\bibitem [{\citenamefont {Mi}\ \emph {et~al.}(2018)\citenamefont {Mi},
  \citenamefont {Benito}, \citenamefont {Putz}, \citenamefont {Zajac},
  \citenamefont {Taylor}, \citenamefont {Burkard},\ and\ \citenamefont
  {Petta}}]{Mi18}%
  \BibitemOpen
  \bibfield  {author} {\bibinfo {author} {\bibfnamefont {X.}~\bibnamefont
  {Mi}}, \bibinfo {author} {\bibfnamefont {M.}~\bibnamefont {Benito}}, \bibinfo
  {author} {\bibfnamefont {S.}~\bibnamefont {Putz}}, \bibinfo {author}
  {\bibfnamefont {D.~M.}\ \bibnamefont {Zajac}}, \bibinfo {author}
  {\bibfnamefont {J.~M.}\ \bibnamefont {Taylor}}, \bibinfo {author}
  {\bibfnamefont {G.}~\bibnamefont {Burkard}}, \ and\ \bibinfo {author}
  {\bibfnamefont {J.~R.}\ \bibnamefont {Petta}},\ }\href {\doibase
  10.1038/nature25769} {\bibfield  {journal} {\bibinfo  {journal} {Nature}\
  }\textbf {\bibinfo {volume} {555}},\ \bibinfo {pages} {599} (\bibinfo {year}
  {2018})}\BibitemShut {NoStop}%
\bibitem [{\citenamefont {Samkharadze}\ \emph {et~al.}(2018)\citenamefont
  {Samkharadze}, \citenamefont {Zheng}, \citenamefont {Kalhor}, \citenamefont
  {Brousse}, \citenamefont {Sammak}, \citenamefont {Mendes}, \citenamefont
  {Blais}, \citenamefont {Scappucci},\ and\ \citenamefont
  {Vandersypen}}]{Samkharadze18}%
  \BibitemOpen
  \bibfield  {author} {\bibinfo {author} {\bibfnamefont {N.}~\bibnamefont
  {Samkharadze}}, \bibinfo {author} {\bibfnamefont {G.}~\bibnamefont {Zheng}},
  \bibinfo {author} {\bibfnamefont {N.}~\bibnamefont {Kalhor}}, \bibinfo
  {author} {\bibfnamefont {D.}~\bibnamefont {Brousse}}, \bibinfo {author}
  {\bibfnamefont {A.}~\bibnamefont {Sammak}}, \bibinfo {author} {\bibfnamefont
  {U.~C.}\ \bibnamefont {Mendes}}, \bibinfo {author} {\bibfnamefont
  {A.}~\bibnamefont {Blais}}, \bibinfo {author} {\bibfnamefont
  {G.}~\bibnamefont {Scappucci}}, \ and\ \bibinfo {author} {\bibfnamefont
  {L.~M.~K.}\ \bibnamefont {Vandersypen}},\ }\href {\doibase
  10.1126/science.aar4054} {\bibfield  {journal} {\bibinfo  {journal}
  {Science}\ }\textbf {\bibinfo {volume} {359}},\ \bibinfo {pages} {1123}
  (\bibinfo {year} {2018})}\BibitemShut {NoStop}%
\bibitem [{\citenamefont {Landig}\ \emph {et~al.}(2018)\citenamefont {Landig},
  \citenamefont {Koski}, \citenamefont {Scarlino}, \citenamefont {Mendes},
  \citenamefont {Blais}, \citenamefont {Reichl}, \citenamefont {Wegscheider},
  \citenamefont {Wallraff}, \citenamefont {Ensslin},\ and\ \citenamefont
  {Ihn}}]{Landig18}%
  \BibitemOpen
  \bibfield  {author} {\bibinfo {author} {\bibfnamefont {A.~J.}\ \bibnamefont
  {Landig}}, \bibinfo {author} {\bibfnamefont {J.~V.}\ \bibnamefont {Koski}},
  \bibinfo {author} {\bibfnamefont {P.}~\bibnamefont {Scarlino}}, \bibinfo
  {author} {\bibfnamefont {U.~C.}\ \bibnamefont {Mendes}}, \bibinfo {author}
  {\bibfnamefont {A.}~\bibnamefont {Blais}}, \bibinfo {author} {\bibfnamefont
  {C.}~\bibnamefont {Reichl}}, \bibinfo {author} {\bibfnamefont
  {W.}~\bibnamefont {Wegscheider}}, \bibinfo {author} {\bibfnamefont
  {A.}~\bibnamefont {Wallraff}}, \bibinfo {author} {\bibfnamefont
  {K.}~\bibnamefont {Ensslin}}, \ and\ \bibinfo {author} {\bibfnamefont
  {T.}~\bibnamefont {Ihn}},\ }\href {\doibase 10.1038/s41586-018-0365-y}
  {\bibfield  {journal} {\bibinfo  {journal} {Nature}\ }\textbf {\bibinfo
  {volume} {560}},\ \bibinfo {pages} {179} (\bibinfo {year}
  {2018})}\BibitemShut {NoStop}%
\bibitem [{\citenamefont {Kawakami}\ \emph {et~al.}(2014)\citenamefont
  {Kawakami}, \citenamefont {Scarlino}, \citenamefont {Ward}, \citenamefont
  {Braakman}, \citenamefont {Savage}, \citenamefont {Lagally}, \citenamefont
  {Friesen}, \citenamefont {Coppersmith}, \citenamefont {Eriksson},\ and\
  \citenamefont {Vandersypen}}]{Kawakami14}%
  \BibitemOpen
  \bibfield  {author} {\bibinfo {author} {\bibfnamefont {E.}~\bibnamefont
  {Kawakami}}, \bibinfo {author} {\bibfnamefont {P.}~\bibnamefont {Scarlino}},
  \bibinfo {author} {\bibfnamefont {D.~R.}\ \bibnamefont {Ward}}, \bibinfo
  {author} {\bibfnamefont {F.~R.}\ \bibnamefont {Braakman}}, \bibinfo {author}
  {\bibfnamefont {D.~E.}\ \bibnamefont {Savage}}, \bibinfo {author}
  {\bibfnamefont {M.~G.}\ \bibnamefont {Lagally}}, \bibinfo {author}
  {\bibfnamefont {M.}~\bibnamefont {Friesen}}, \bibinfo {author} {\bibfnamefont
  {S.~N.}\ \bibnamefont {Coppersmith}}, \bibinfo {author} {\bibfnamefont
  {M.~A.}\ \bibnamefont {Eriksson}}, \ and\ \bibinfo {author} {\bibfnamefont
  {L.~M.~K.}\ \bibnamefont {Vandersypen}},\ }\href
  {https://doi.org/10.1038/nnano.2014.153} {\bibfield  {journal} {\bibinfo
  {journal} {Nature Nanotechnology}\ }\textbf {\bibinfo {volume} {9}},\
  \bibinfo {pages} {666} (\bibinfo {year} {2014})}\BibitemShut {NoStop}%
\bibitem [{\citenamefont {Mi}\ \emph {et~al.}(2017)\citenamefont {Mi},
  \citenamefont {Cady}, \citenamefont {Zajac}, \citenamefont {Deelman},\ and\
  \citenamefont {Petta}}]{Mi17}%
  \BibitemOpen
  \bibfield  {author} {\bibinfo {author} {\bibfnamefont {X.}~\bibnamefont
  {Mi}}, \bibinfo {author} {\bibfnamefont {J.~V.}\ \bibnamefont {Cady}},
  \bibinfo {author} {\bibfnamefont {D.~M.}\ \bibnamefont {Zajac}}, \bibinfo
  {author} {\bibfnamefont {P.~W.}\ \bibnamefont {Deelman}}, \ and\ \bibinfo
  {author} {\bibfnamefont {J.~R.}\ \bibnamefont {Petta}},\ }\href {\doibase
  10.1126/science.aal2469} {\bibfield  {journal} {\bibinfo  {journal}
  {Science}\ }\textbf {\bibinfo {volume} {355}},\ \bibinfo {pages} {156}
  (\bibinfo {year} {2017})}\BibitemShut {NoStop}%
\bibitem [{\citenamefont {Stockklauser}\ \emph {et~al.}(2017)\citenamefont
  {Stockklauser}, \citenamefont {Scarlino}, \citenamefont {Koski},
  \citenamefont {Gasparinetti}, \citenamefont {Andersen}, \citenamefont
  {Reichl}, \citenamefont {Wegscheider}, \citenamefont {Ihn}, \citenamefont
  {Ensslin},\ and\ \citenamefont {Wallraff}}]{Stockklauser17}%
  \BibitemOpen
  \bibfield  {author} {\bibinfo {author} {\bibfnamefont {A.}~\bibnamefont
  {Stockklauser}}, \bibinfo {author} {\bibfnamefont {P.}~\bibnamefont
  {Scarlino}}, \bibinfo {author} {\bibfnamefont {J.~V.}\ \bibnamefont {Koski}},
  \bibinfo {author} {\bibfnamefont {S.}~\bibnamefont {Gasparinetti}}, \bibinfo
  {author} {\bibfnamefont {C.~K.}\ \bibnamefont {Andersen}}, \bibinfo {author}
  {\bibfnamefont {C.}~\bibnamefont {Reichl}}, \bibinfo {author} {\bibfnamefont
  {W.}~\bibnamefont {Wegscheider}}, \bibinfo {author} {\bibfnamefont
  {T.}~\bibnamefont {Ihn}}, \bibinfo {author} {\bibfnamefont {K.}~\bibnamefont
  {Ensslin}}, \ and\ \bibinfo {author} {\bibfnamefont {A.}~\bibnamefont
  {Wallraff}},\ }\href {\doibase 10.1103/PhysRevX.7.011030} {\bibfield
  {journal} {\bibinfo  {journal} {Phys. Rev. X}\ }\textbf {\bibinfo {volume}
  {7}},\ \bibinfo {pages} {011030} (\bibinfo {year} {2017})}\BibitemShut
  {NoStop}%
\bibitem [{\citenamefont {van Woerkom}\ \emph {et~al.}(2018)\citenamefont {van
  Woerkom}, \citenamefont {Scarlino}, \citenamefont {Ungerer}, \citenamefont
  {M\"{u}ller}, \citenamefont {Koski}, \citenamefont {Landig}, \citenamefont
  {Reichl}, \citenamefont {Wegscheider}, \citenamefont {Ihn}, \citenamefont
  {Ensslin},\ and\ \citenamefont {Wallraff}}]{Woerkom18}%
  \BibitemOpen
  \bibfield  {author} {\bibinfo {author} {\bibfnamefont {D.~J.}\ \bibnamefont
  {van Woerkom}}, \bibinfo {author} {\bibfnamefont {P.}~\bibnamefont
  {Scarlino}}, \bibinfo {author} {\bibfnamefont {J.~H.}\ \bibnamefont
  {Ungerer}}, \bibinfo {author} {\bibfnamefont {C.}~\bibnamefont {M\"{u}ller}},
  \bibinfo {author} {\bibfnamefont {J.~V.}\ \bibnamefont {Koski}}, \bibinfo
  {author} {\bibfnamefont {A.~J.}\ \bibnamefont {Landig}}, \bibinfo {author}
  {\bibfnamefont {C.}~\bibnamefont {Reichl}}, \bibinfo {author} {\bibfnamefont
  {W.}~\bibnamefont {Wegscheider}}, \bibinfo {author} {\bibfnamefont
  {T.}~\bibnamefont {Ihn}}, \bibinfo {author} {\bibfnamefont {K.}~\bibnamefont
  {Ensslin}}, \ and\ \bibinfo {author} {\bibfnamefont {A.}~\bibnamefont
  {Wallraff}},\ }\href {\doibase 10.1103/PhysRevX.8.041018} {\bibfield
  {journal} {\bibinfo  {journal} {Phys. Rev. X}\ }\textbf {\bibinfo {volume}
  {8}},\ \bibinfo {pages} {041018} (\bibinfo {year} {2018})}\BibitemShut
  {NoStop}%
\bibitem [{\citenamefont {Cottet}\ and\ \citenamefont
  {Kontos}(2010)}]{Cottet10}%
  \BibitemOpen
  \bibfield  {author} {\bibinfo {author} {\bibfnamefont {A.}~\bibnamefont
  {Cottet}}\ and\ \bibinfo {author} {\bibfnamefont {T.}~\bibnamefont
  {Kontos}},\ }\href {\doibase 10.1103/PhysRevLett.105.160502} {\bibfield
  {journal} {\bibinfo  {journal} {Phys. Rev. Lett.}\ }\textbf {\bibinfo
  {volume} {105}},\ \bibinfo {pages} {160502} (\bibinfo {year}
  {2010})}\BibitemShut {NoStop}%
\bibitem [{\citenamefont {Hu}\ \emph {et~al.}(2012)\citenamefont {Hu},
  \citenamefont {Liu},\ and\ \citenamefont {Nori}}]{Hu12}%
  \BibitemOpen
  \bibfield  {author} {\bibinfo {author} {\bibfnamefont {X.}~\bibnamefont
  {Hu}}, \bibinfo {author} {\bibfnamefont {Y.-x.}\ \bibnamefont {Liu}}, \ and\
  \bibinfo {author} {\bibfnamefont {F.}~\bibnamefont {Nori}},\ }\href {\doibase
  10.1103/PhysRevB.86.035314} {\bibfield  {journal} {\bibinfo  {journal} {Phys.
  Rev. B}\ }\textbf {\bibinfo {volume} {86}},\ \bibinfo {pages} {035314}
  (\bibinfo {year} {2012})}\BibitemShut {NoStop}%
\bibitem [{\citenamefont {Wu}\ \emph {et~al.}(2018)\citenamefont {Wu},
  \citenamefont {Cheng}, \citenamefont {Yu},\ and\ \citenamefont
  {Wang}}]{Wu18}%
  \BibitemOpen
  \bibfield  {author} {\bibinfo {author} {\bibfnamefont {S.}~\bibnamefont
  {Wu}}, \bibinfo {author} {\bibfnamefont {L.}~\bibnamefont {Cheng}}, \bibinfo
  {author} {\bibfnamefont {H.}~\bibnamefont {Yu}}, \ and\ \bibinfo {author}
  {\bibfnamefont {Q.}~\bibnamefont {Wang}},\ }\href {\doibase
  https://doi.org/10.1016/j.physleta.2018.05.005} {\bibfield  {journal}
  {\bibinfo  {journal} {Physics Letters A}\ }\textbf {\bibinfo {volume}
  {382}},\ \bibinfo {pages} {1922} (\bibinfo {year} {2018})}\BibitemShut
  {NoStop}%
\bibitem [{\citenamefont {DiVincenzo}(2000)}]{DiVincenzo00}%
  \BibitemOpen
  \bibfield  {author} {\bibinfo {author} {\bibfnamefont {D.~P.}\ \bibnamefont
  {DiVincenzo}},\ }\href {\doibase
  10.1002/1521-3978(200009)48:9/11<771::AID-PROP771>3.0.CO;2-E} {\bibfield
  {journal} {\bibinfo  {journal} {Fortschritte der Physik}\ }\textbf {\bibinfo
  {volume} {48}},\ \bibinfo {pages} {771} (\bibinfo {year} {2000})}\BibitemShut
  {NoStop}%
\bibitem [{\citenamefont {Beaudoin}\ \emph {et~al.}(2016)\citenamefont
  {Beaudoin}, \citenamefont {Lachance-Quirion}, \citenamefont {Coish},\ and\
  \citenamefont {Pioro-Ladri\`{e}re}}]{Beaudoin16}%
  \BibitemOpen
  \bibfield  {author} {\bibinfo {author} {\bibfnamefont {F.}~\bibnamefont
  {Beaudoin}}, \bibinfo {author} {\bibfnamefont {D.}~\bibnamefont
  {Lachance-Quirion}}, \bibinfo {author} {\bibfnamefont {W.~A.}\ \bibnamefont
  {Coish}}, \ and\ \bibinfo {author} {\bibfnamefont {M.}~\bibnamefont
  {Pioro-Ladri\`{e}re}},\ }\href {\doibase 10.1088/0957-4484/27/46/464003}
  {\bibfield  {journal} {\bibinfo  {journal} {Nanotechnology}\ }\textbf
  {\bibinfo {volume} {27}},\ \bibinfo {pages} {464003} (\bibinfo {year}
  {2016})}\BibitemShut {NoStop}%
\bibitem [{\citenamefont {Hanson}\ \emph {et~al.}(2007)\citenamefont {Hanson},
  \citenamefont {Kouwenhoven}, \citenamefont {Petta}, \citenamefont {Tarucha},\
  and\ \citenamefont {Vandersypen}}]{Hanson07}%
  \BibitemOpen
  \bibfield  {author} {\bibinfo {author} {\bibfnamefont {R.}~\bibnamefont
  {Hanson}}, \bibinfo {author} {\bibfnamefont {L.~P.}\ \bibnamefont
  {Kouwenhoven}}, \bibinfo {author} {\bibfnamefont {J.~R.}\ \bibnamefont
  {Petta}}, \bibinfo {author} {\bibfnamefont {S.}~\bibnamefont {Tarucha}}, \
  and\ \bibinfo {author} {\bibfnamefont {L.~M.~K.}\ \bibnamefont
  {Vandersypen}},\ }\href {\doibase 10.1103/RevModPhys.79.1217} {\bibfield
  {journal} {\bibinfo  {journal} {Rev. Mod. Phys.}\ }\textbf {\bibinfo {volume}
  {79}},\ \bibinfo {pages} {1217} (\bibinfo {year} {2007})}\BibitemShut
  {NoStop}%
\bibitem [{\citenamefont {Schrieffer}\ and\ \citenamefont
  {Wolff}(1966)}]{Schrieffer66}%
  \BibitemOpen
  \bibfield  {author} {\bibinfo {author} {\bibfnamefont {J.~R.}\ \bibnamefont
  {Schrieffer}}\ and\ \bibinfo {author} {\bibfnamefont {P.~A.}\ \bibnamefont
  {Wolff}},\ }\href {\doibase 10.1103/PhysRev.149.491} {\bibfield  {journal}
  {\bibinfo  {journal} {Phys. Rev.}\ }\textbf {\bibinfo {volume} {149}},\
  \bibinfo {pages} {491} (\bibinfo {year} {1966})}\BibitemShut {NoStop}%
\bibitem [{Sup()}]{Supplement}%
  \BibitemOpen
  \href@noop {} {}\bibinfo {note} {See Supplemental Material at [URL will be
  inserted by publisher] for a full derivation of the effective
  Hamiltonian.}\BibitemShut {Stop}%
\bibitem [{\citenamefont {McKay}\ \emph {et~al.}(2016)\citenamefont {McKay},
  \citenamefont {Filipp}, \citenamefont {Mezzacapo}, \citenamefont {Magesan},
  \citenamefont {Chow},\ and\ \citenamefont {Gambetta}}]{McKay16}%
  \BibitemOpen
  \bibfield  {author} {\bibinfo {author} {\bibfnamefont {D.~C.}\ \bibnamefont
  {McKay}}, \bibinfo {author} {\bibfnamefont {S.}~\bibnamefont {Filipp}},
  \bibinfo {author} {\bibfnamefont {A.}~\bibnamefont {Mezzacapo}}, \bibinfo
  {author} {\bibfnamefont {E.}~\bibnamefont {Magesan}}, \bibinfo {author}
  {\bibfnamefont {J.~M.}\ \bibnamefont {Chow}}, \ and\ \bibinfo {author}
  {\bibfnamefont {J.~M.}\ \bibnamefont {Gambetta}},\ }\href {\doibase
  10.1103/PhysRevApplied.6.064007} {\bibfield  {journal} {\bibinfo  {journal}
  {Phys. Rev. Applied}\ }\textbf {\bibinfo {volume} {6}},\ \bibinfo {pages}
  {064007} (\bibinfo {year} {2016})}\BibitemShut {NoStop}%
\bibitem [{\citenamefont {Rezakhani}(2004)}]{Rezakhani04}%
  \BibitemOpen
  \bibfield  {author} {\bibinfo {author} {\bibfnamefont {A.~T.}\ \bibnamefont
  {Rezakhani}},\ }\href {\doibase 10.1103/PhysRevA.70.052313} {\bibfield
  {journal} {\bibinfo  {journal} {Phys. Rev. A}\ }\textbf {\bibinfo {volume}
  {70}},\ \bibinfo {pages} {052313} (\bibinfo {year} {2004})}\BibitemShut
  {NoStop}%
\bibitem [{\citenamefont {Srinivasa}\ \emph {et~al.}(2016)\citenamefont
  {Srinivasa}, \citenamefont {Taylor},\ and\ \citenamefont
  {Tahan}}]{Srinivasa16}%
  \BibitemOpen
  \bibfield  {author} {\bibinfo {author} {\bibfnamefont {V.}~\bibnamefont
  {Srinivasa}}, \bibinfo {author} {\bibfnamefont {J.~M.}\ \bibnamefont
  {Taylor}}, \ and\ \bibinfo {author} {\bibfnamefont {C.}~\bibnamefont
  {Tahan}},\ }\href {\doibase 10.1103/PhysRevB.94.205421} {\bibfield  {journal}
  {\bibinfo  {journal} {Phys. Rev. B}\ }\textbf {\bibinfo {volume} {94}},\
  \bibinfo {pages} {205421} (\bibinfo {year} {2016})}\BibitemShut {NoStop}%
\bibitem [{\citenamefont {Clarence}\ and\ \citenamefont
  {Howard}(1932)}]{Zener32}%
  \BibitemOpen
  \bibfield  {author} {\bibinfo {author} {\bibfnamefont {Z.}~\bibnamefont
  {Clarence}}\ and\ \bibinfo {author} {\bibfnamefont {F.~R.}\ \bibnamefont
  {Howard}},\ }\href {\doibase 10.1098/rspa.1932.0165} {\bibfield  {journal}
  {\bibinfo  {journal} {Proceedings of the Royal Society of London. Series A}\
  }\textbf {\bibinfo {volume} {137}},\ \bibinfo {pages} {696} (\bibinfo {year}
  {1932})}\BibitemShut {NoStop}%
\bibitem [{\citenamefont {Petta}\ \emph {et~al.}(2005)\citenamefont {Petta},
  \citenamefont {Johnson}, \citenamefont {Taylor}, \citenamefont {Laird},
  \citenamefont {Yacoby}, \citenamefont {Lukin}, \citenamefont {Marcus},
  \citenamefont {Hanson},\ and\ \citenamefont {Gossard}}]{Petta05}%
  \BibitemOpen
  \bibfield  {author} {\bibinfo {author} {\bibfnamefont {J.~R.}\ \bibnamefont
  {Petta}}, \bibinfo {author} {\bibfnamefont {A.~C.}\ \bibnamefont {Johnson}},
  \bibinfo {author} {\bibfnamefont {J.~M.}\ \bibnamefont {Taylor}}, \bibinfo
  {author} {\bibfnamefont {E.~A.}\ \bibnamefont {Laird}}, \bibinfo {author}
  {\bibfnamefont {A.}~\bibnamefont {Yacoby}}, \bibinfo {author} {\bibfnamefont
  {M.~D.}\ \bibnamefont {Lukin}}, \bibinfo {author} {\bibfnamefont {C.~M.}\
  \bibnamefont {Marcus}}, \bibinfo {author} {\bibfnamefont {M.~P.}\
  \bibnamefont {Hanson}}, \ and\ \bibinfo {author} {\bibfnamefont {A.~C.}\
  \bibnamefont {Gossard}},\ }\href {\doibase 10.1126/science.1116955}
  {\bibfield  {journal} {\bibinfo  {journal} {Science}\ }\textbf {\bibinfo
  {volume} {309}},\ \bibinfo {pages} {2180} (\bibinfo {year}
  {2005})}\BibitemShut {NoStop}%
\bibitem [{\citenamefont {Bravyi}\ \emph {et~al.}(2011)\citenamefont {Bravyi},
  \citenamefont {DiVincenzo},\ and\ \citenamefont {Loss}}]{Bravyi11}%
  \BibitemOpen
  \bibfield  {author} {\bibinfo {author} {\bibfnamefont {S.}~\bibnamefont
  {Bravyi}}, \bibinfo {author} {\bibfnamefont {D.~P.}\ \bibnamefont
  {DiVincenzo}}, \ and\ \bibinfo {author} {\bibfnamefont {D.}~\bibnamefont
  {Loss}},\ }\href {\doibase 10.1016/j.aop.2011.06.004} {\bibfield  {journal}
  {\bibinfo  {journal} {Annals of Physics}\ }\textbf {\bibinfo {volume}
  {326}},\ \bibinfo {pages} {2793} (\bibinfo {year} {2011})}\BibitemShut
  {NoStop}%
\bibitem [{\citenamefont {Nielsen}(2002)}]{Nielsen02}%
  \BibitemOpen
  \bibfield  {author} {\bibinfo {author} {\bibfnamefont {M.~A.}\ \bibnamefont
  {Nielsen}},\ }\href {\doibase 10.1016/S0375-9601(02)01272-0} {\bibfield
  {journal} {\bibinfo  {journal} {Physics Letters A}\ }\textbf {\bibinfo
  {volume} {303}},\ \bibinfo {pages} {249} (\bibinfo {year}
  {2002})}\BibitemShut {NoStop}%
\bibitem [{\citenamefont {Benito}\ \emph {et~al.}(2019)\citenamefont {Benito},
  \citenamefont {Petta},\ and\ \citenamefont {Burkard}}]{Benito19}%
  \BibitemOpen
  \bibfield  {author} {\bibinfo {author} {\bibfnamefont {M.}~\bibnamefont
  {Benito}}, \bibinfo {author} {\bibfnamefont {J.~R.}\ \bibnamefont {Petta}}, \
  and\ \bibinfo {author} {\bibfnamefont {G.}~\bibnamefont {Burkard}},\
  }\href@noop {} {\bibfield  {journal} {\bibinfo  {journal} {arXiv:1902.07649}\
  } (\bibinfo {year} {2019})}\BibitemShut {NoStop}%
\end{thebibliography}%


\begin{thebibliography}{2}%
\makeatletter
\providecommand \@ifxundefined [1]{%
 \@ifx{#1\undefined}
}%
\providecommand \@ifnum [1]{%
 \ifnum #1\expandafter \@firstoftwo
 \else \expandafter \@secondoftwo
 \fi
}%
\providecommand \@ifx [1]{%
 \ifx #1\expandafter \@firstoftwo
 \else \expandafter \@secondoftwo
 \fi
}%
\providecommand \natexlab [1]{#1}%
\providecommand \enquote  [1]{``#1''}%
\providecommand \bibnamefont  [1]{#1}%
\providecommand \bibfnamefont [1]{#1}%
\providecommand \citenamefont [1]{#1}%
\providecommand \href@noop [0]{\@secondoftwo}%
\providecommand \href [0]{\begingroup \@sanitize@url \@href}%
\providecommand \@href[1]{\@@startlink{#1}\@@href}%
\providecommand \@@href[1]{\endgroup#1\@@endlink}%
\providecommand \@sanitize@url [0]{\catcode `\\12\catcode `\$12\catcode
  `\&12\catcode `\#12\catcode `\^12\catcode `\_12\catcode `\%12\relax}%
\providecommand \@@startlink[1]{}%
\providecommand \@@endlink[0]{}%
\providecommand \url  [0]{\begingroup\@sanitize@url \@url }%
\providecommand \@url [1]{\endgroup\@href {#1}{\urlprefix }}%
\providecommand \urlprefix  [0]{URL }%
\providecommand \Eprint [0]{\href }%
\providecommand \doibase [0]{http://dx.doi.org/}%
\providecommand \selectlanguage [0]{\@gobble}%
\providecommand \bibinfo  [0]{\@secondoftwo}%
\providecommand \bibfield  [0]{\@secondoftwo}%
\providecommand \translation [1]{[#1]}%
\providecommand \BibitemOpen [0]{}%
\providecommand \bibitemStop [0]{}%
\providecommand \bibitemNoStop [0]{.\EOS\space}%
\providecommand \EOS [0]{\spacefactor3000\relax}%
\providecommand \BibitemShut  [1]{\csname bibitem#1\endcsname}%
\let\auto@bib@innerbib\@empty
\bibitem [{\citenamefont {Bravyi}\ \emph {et~al.}(2011)\citenamefont {Bravyi},
  \citenamefont {DiVincenzo},\ and\ \citenamefont {Loss}}]{Bravyi11}%
  \BibitemOpen
  \bibfield  {author} {\bibinfo {author} {\bibfnamefont {S.}~\bibnamefont
  {Bravyi}}, \bibinfo {author} {\bibfnamefont {D.~P.}\ \bibnamefont
  {DiVincenzo}}, \ and\ \bibinfo {author} {\bibfnamefont {D.}~\bibnamefont
  {Loss}},\ }\href {\doibase 10.1016/j.aop.2011.06.004} {\bibfield  {journal}
  {\bibinfo  {journal} {Annals of Physics}\ }\textbf {\bibinfo {volume}
  {326}},\ \bibinfo {pages} {2793} (\bibinfo {year} {2011})}\BibitemShut
  {NoStop}%
\bibitem [{\citenamefont {Beaudoin}\ \emph {et~al.}(2016)\citenamefont
  {Beaudoin}, \citenamefont {Lachance-Quirion}, \citenamefont {Coish},\ and\
  \citenamefont {Pioro-Ladri\`{e}re}}]{Beaudoin16}%
  \BibitemOpen
  \bibfield  {author} {\bibinfo {author} {\bibfnamefont {F.}~\bibnamefont
  {Beaudoin}}, \bibinfo {author} {\bibfnamefont {D.}~\bibnamefont
  {Lachance-Quirion}}, \bibinfo {author} {\bibfnamefont {W.~A.}\ \bibnamefont
  {Coish}}, \ and\ \bibinfo {author} {\bibfnamefont {M.}~\bibnamefont
  {Pioro-Ladri\`{e}re}},\ }\href {\doibase 10.1088/0957-4484/27/46/464003}
  {\bibfield  {journal} {\bibinfo  {journal} {Nanotechnology}\ }\textbf
  {\bibinfo {volume} {27}},\ \bibinfo {pages} {464003} (\bibinfo {year}
  {2016})}\BibitemShut {NoStop}%
\end{thebibliography}%

\end{document}


\title{Supplemental material to ``Long-distance entangling gates between quantum dot spins mediated by a superconducting resonator"}

\author{Ada Warren}
\author{Edwin Barnes}
\author{Sophia E. Economou}
\affiliation{Department of Physics, Virginia Tech, Blacksburg VA 24061, USA}

\maketitle
\onecolumngrid

In these supplemental materials, we present a derivation of the effective $N$-spin Hamiltonian presented in the main text from the original $N$-DQD Hamiltonian. We use the $\{\ket{+}_i, \ket{-}_i\}$ orbital eigenbasis introduced into the main text, and we start with the Hamiltonian $H_N = H_0 + V$ where
\begin{equation}
H_0 = \hbar \omega_r a^\dag a + \sum_{i = 1}^N \qty(\frac{1}{2} \hbar \omega_{a i} \tau_{z i} + \frac{1}{2} \hbar \omega_{z i} \sigma_{z i}),
\end{equation}
and
\begin{align}
V =& \sum_{i = 1}^N \qty(g_{x i}\sigma_{x i} + g_{z i} \sigma_{z i})\qty(\cos(\theta_i) \tau_{z i} - \sin(\theta_i) \tau_{x i}) \\
	&+ (a^\dag + a)\sum_{i = 1}^N g_{AC i} \qty(1 - \cos(\theta_i)\tau_{z i} + \sin(\theta_i) \tau_{x i}). \nonumber
\end{align}
We also introduce the superoperator
\begin{equation}
\mathcal{L}(X) = \sum_{i,j}\dyad{i}{i}\frac{X}{E_i - E_j}\dyad{j}{j},
\end{equation}
where the sum is taken over all eigenstates $\ket{i}$ and $\ket{j}$ of $H_0$ and where $H_0 \ket{i} = E_i \ket{i}$. Note that the action of this superoperator is only well-defined on operators which are purely off-diagonal in the $H_0$ eigenbasis.

We now wish to apply the Schrieffer-Wolff transformation $e^{S}$, where $S$ is an anti-unitary operator such that $e^{-S} H_N e^{S}$ contains no block off-diagonal terms which couple the ground and excited orbital states (i.e. no $\tau_x$ or $\tau_y$) to leading order \cite{Bravyi11}. We partition the perturbation $V$ into block diagonal and block off-diagonal terms $V = V_d + V_{od}$ where
\begin{equation}
V_d = \sum_{i = 1}^N \qty(\qty(g_{x i}\sigma_{x i} + g_{z i} \sigma_{z i})\cos(\theta_i) \tau_{z i} + g_{AC i} (a^\dag + a) \qty(1 - \cos(\theta_i)\tau_{z i})),
\end{equation}
and
\begin{equation}
V_{od} = \sum_{i = 1}^N \sin(\theta_i) \qty(g_{AC i} \qty(a^\dag + a) - \qty(g_{x i}\sigma_{x i} + g_{z i} \sigma_{z i}) )\tau_{x i}.
\end{equation}

Noting that $S$ will be at least first order in $V$, we apply the transformation to second order in $V$ to obtain
\begin{align}
e^{-S} H_N e^{S} \approx& H_0 + V_d + V_{od} + \comm{S}{H_0} \\
	& + \comm{S}{V_d} + \comm{S}{V_{od}} + \frac{1}{2}\comm{S}{\comm{S}{H_0}} \nonumber \\
	& + \mathcal{O}(V^3). \nonumber
\end{align}
We demand that all block off-diagonal terms in this expression vanish to first order in $V$. This is clearly satisfied if $\comm{S}{H_0} + V_{od} = 0$, and it is easy to verify that this is the case if $S = \mathcal{L}(V_{od})$ \cite{Bravyi11}. Evaluating, we obtain
\begin{align}
S = \frac{i}{\hbar} \sum_{j = 1}^N& \sin(\theta_j) ( \frac{g_{AC j}}{\omega_{a j}^2 - \omega_r^2} \qty(\omega_{a j}\qty(a^\dag + a) \tau_{y j} - \omega_r \qty(\frac{a^\dag - a}{i}) \tau_{x j}) \\
	& + \frac{g_{x j}}{\omega_{a j}^2 - \omega_{z j}^2} \qty(\omega_{z j} \sigma_{y j} \tau_{x j} - \omega_{a j} \sigma_{x j} \tau_{y j}) - \frac{g_{z j}}{\omega_{a j}} \sigma_{z j} \tau_{y j} ). \nonumber
\end{align}

Now we define $P$ to be the asymmetric projector onto the orbital ground subspace
\begin{equation}
P = \sum_{s_1,...,s_N,n}\dyad{\{s_1,...,s_N\},\{-,...,-\},n}{\{s_1,...,s_N\},n}.
\end{equation}
We take our transformed Hamiltonian and project onto the subspace to arrive at an effective spin-resonator Hamiltonian. Dropping additive constants,
\begin{align}
\label{eq:firsteff}
H'_N &= P^\dag e^{-S} H_N e^S P \\
	&\approx P^\dag \qty(H_0 + V_d + \frac{1}{2}\comm{S}{V_{od}}) P \nonumber \\
	&= H'_0 + V', \nonumber
\end{align}
where
\begin{align}
H'_0 &= \hbar \omega'_r a^\dag a + \sum_{j = 1}^N (\frac{1}{2}\hbar\omega_{z j} \sigma_{z j} - \cos(\theta_j) \qty(g_{x j}\sigma_{x j} + g_{z j} \sigma_{z j}) \\
	& \quad + \sin[2](\theta_j) \frac{g_{x j} \omega_{z j}}{\hbar\qty(\omega_{a j}^2 - \omega_{z j}^2)} \qty(g_{z j}\sigma_{x j} - g_{x j} \sigma_{z j}) ), \nonumber
\end{align}
and
\begin{align}
V' &= \sum_{j = 1}^N (\qty(g'_{x j} \sigma_{x j} + g'_{z j} \sigma_{z j} + g_{AC j}\qty(1 + \cos(\theta_j))) \qty(a^\dag + a) \\
	& \quad - \sin[2](\theta_j) \frac{\omega_{a j} g_{AC j}^2}{\hbar\qty(\omega_{a j}^2 - \omega_r^2)} \qty(a^{\dag 2} + a^2) ). \nonumber
\end{align}
We define the primed constants
\begin{align*}
&\omega'_r = \omega_r - 2 \sum_{i=1}^N \frac{g_{AC i}^2}{\hbar^2} \sin[2](\theta_i) \frac{\omega_{a i}}{\omega_{a i}^2 - \omega_r^2}, \\
&g'_{x i} = g_{x i} \frac{g_{AC i}}{\hbar} \sin[2](\theta_i) \omega_{a i} \qty(\frac{1}{\omega_{a i}^2 - \omega_{z i}^2} + \frac{1}{\omega_{a i}^2 - \omega_r^2}), \\
&g'_{z i} = g_{z i} \frac{g_{AC i}}{\hbar} \sin[2](\theta_i) \omega_{a i} \qty(\frac{1}{\omega_{a i}^2} + \frac{1}{\omega_{a i}^2 - \omega_r^2}).
\end{align*}
It is convenient for the next step to work in a basis where $H'_0$ is diagonal. This can be achieved with a $y$-rotation of the spin bases. This rotation simplifies the resulting expressions, but is sufficiently small that we can, to good approximation, treat the result as acting on the original spin basis. Performing the rotation yields
\begin{equation}
H'_0 = \omega'_r a^\dag a + \sum_{j = 1}^N \frac{1}{2} \hbar \omega'_{z j} \sigma'_{z j},
\end{equation}
where we define
\begin{equation*}
\omega'_{z i} = \omega_{z i} - 2 \cos(\theta_i) \frac{g_{z i}}{\hbar} + 2  \omega_{z i} \qty(\frac{\cos[2](\theta_i)}{\omega_{z i}^2} - \frac{\sin[2](\theta_i)}{\omega_{a i}^2 - \omega_{z i}^2}) \frac{g_{x i}^2}{\hbar^2},
\end{equation*}
and where $\vec{\sigma'}$ are the Pauli operators in this new spin basis. This rotation leaves the form of $V'$ unchanged to second order in $V$.

At this point, we have derived both resonator energy corrections $\omega'_r$ as well as effective spin-resonator couplings \cite{Beaudoin16}. Our goal, however, is to obtain an effective spin-spin Hamiltonian. To this end, we apply another Schrieffer-Wolff transformation $e^{S'}$, this time to eliminate the couplings between the ground and excited resonator states. We proceed as before, working this time to second order in $V'$. This approach is inconsistent mathematically, as we are ultimately after terms which are fourth order in $V$, but we have already neglected some terms which were third and fourth order in $V$. The ultimate justification of this step is the agreement with numerical results presented in the main text.

We once again define a superoperator
\begin{equation}
\mathcal{L}'(X) = \sum_{i,j}\dyad{i}{i}\frac{X}{E'_i - E'_j}\dyad{j}{j},
\end{equation}
where now the sum is taken over the eigenstates of $H'_0$. Noting that this time, the perturbation $V'$ contains no block-diagonal terms, it is easy to show that the correct transformation is
\begin{align}
S' =& \mathcal{L}'(V') \\
	=& \frac{i}{\hbar} \sum_{j = 1}^N (\frac{g'_{x j}}{\omega^{'2}_r - \omega^{'2}_{z j}}\qty(\omega'_r \sigma'_{x j}\qty(\frac{a^\dag - a}{i}) - \omega'_{z j} \sigma'_{y j} \qty(a^\dag + a)) \nonumber\\
	& \quad + \frac{1}{\omega'_r}\qty(\frac{a^\dag - a}{i})\qty(g'_{z j} \sigma'_{z j} + g_{AC j}\qty(1 - \cos(\theta_j))) \nonumber\\
	& \quad - \sin[2](\theta_j) \frac{\omega_{a j} g_{AC j}^2}{2 \omega'_r \qty(\omega_{a j}^2 - \omega_r^2)}\qty(\frac{a^{\dag 2} - a^2}{i})). \nonumber
\end{align}

Now, let $P'$ be the asymmetric projector onto the low-energy subspace
\begin{equation}
P' = \sum_{s_1,...,s_N} \dyad{\{s_1,...,s_N\},0}{\{s_1,...,s_N\}}.
\end{equation}
Again, we take our transformed Hamiltonian, project onto the low-energy subspace, and drop any additive constants to arrive at
\begin{align}
H''_N =& P^{'\dag} \qty(H'_0 + \frac{1}{2}\comm{S'}{V'}) P' \\
	=& \sum_{i = 1}^N \qty( \frac{1}{2} \hbar \omega'_{z i} \sigma'_{z i} + \frac{g'_{x i} \omega'_{z i}}{\hbar\qty(\omega^{'2}_r - \omega^{'2}_{z i})} \qty(g'_{z i} \sigma'_{x i} - g'_{x i} \sigma'_{z i}) - \qty(\qty(\frac{1}{\omega^{'2}_r - \omega^{'2}_{z i}} + \frac{1}{\omega^{'2}_r})\omega'_r g'_{x i}\sigma'_{x i} + \frac{2 g'_{z i}}{\omega'_r} \sigma'_{z i})\frac{\Sigma}{\hbar} ) \nonumber \\
	&- \sum_{i \neq j} \frac{\omega'_r}{\hbar} \qty(\frac{g'_{x j}}{\omega_{r}^{'2} - \omega_{z j}^{'2}}\sigma'_{x j} + \frac{g'_{z j}}{\omega_{r}^{'2}}\sigma'_{z j})\qty(g'_{x i} \sigma'_{x i} + g'_{z i} \sigma'_{z i}), \nonumber
\end{align}
where we define
\begin{equation*}
\Sigma = \sum_{j=1}^N g_{AC j}\qty(1 + \cos(\theta_j)).
\end{equation*}
This is our effective spin-spin Hamiltonian. It is once again desirable, however, to work in a basis in which the single-spin operators are diagonal. If we perform another very small spin basis rotation to diagonalize the first sum, working to second order in $V'$, we finally obtain the desired effective $N$-spin Hamiltonian
\begin{equation}
H''_N = \sum_{i = 1}^N \frac{1}{2} \hbar \omega''_{z i} \sigma''_{z i} - \sum_{i \neq j} \frac{\omega'_r}{\hbar} \qty(\frac{g'_{x j}}{\omega_{r}^{'2} - \omega_{z j}^{'2}}\sigma''_{x j} + \frac{g'_{z j}}{\omega_{r}^{'2}}\sigma''_{z j})\qty(g'_{x i} \sigma''_{x i} + g'_{z i} \sigma''_{z i}),
\end{equation}
where $\vec{\sigma''}$ are the Pauli operators in the new spin basis, and where we have defined the dressed spin splittings
\begin{equation*}
\omega''_{z i} = \omega'_{z i} - 2 \frac{g_{x i}^{'2} \omega'_{z i}}{\hbar^2 \qty(\omega_r^{'2} - \omega_{z i}^{'2})} - 4 \frac{g'_{z i}}{\hbar^2 \omega'_r}\Sigma.
\end{equation*}

\bibliography{refs}{}